\documentclass[prl, twocolumn, showpacs]{revtex4}

\usepackage{amssymb}
\usepackage{amsmath}
\usepackage{graphicx}
\makeatletter

\begin{document}
   \DeclareGraphicsExtensions{.pdf,.gif,.jpg.eps}

\title{Andreev and Single Particle Tunneling Spectroscopies in Underdoped Cuprates}
\author{Kai-Yu Yang$^{1,2}$, Kun Huang$^{2}$, Wei-Qiang Chen$^{2}$, T. M. Rice$^{1,2,3}$, and Fu-Chun Zhang$^{2}$}

\address{
$^{1}$ Institut f$\ddot{u}$r Theoretische Physik, ETH Z$\ddot{u}$rich,CH-8093 Z$\ddot{u}$rich, Switzerland \\
$^{2}$ Center for Theoretical and Computational Physics and
Department of Physics, The University of Hong Kong, Hong Kong SAR,
China \\
$^{3}$ Condensed Matter Physics and Materials Science Department, Brookhaven National Laboratory, Upton, NY 11973, USA}

\begin{abstract}
We study tunneling spectroscopy between a normal metal and
underdoped cuprate superconductor modeled by a phenomenological
theory in which the pseudogap is a precursor to the undoped Mott
insulator. In the transparent tunneling limit, the spectra show a
small energy gap associated with Andreev reflection. In the
Giaever limit, the spectra show a large energy gap associated with
single particle tunneling. Our theory semi-quantitatively describes
the two gap behavior observed in tunneling experiments.
\end{abstract}

\date{\today}
\maketitle

  It has long been argued that the highly anomalous pseudogap phase of underdoped (UD)
  cuprates holds the key to the physics of high $T_{c}$
  superconductors \cite{Lee-Wen-RMP, Ogata-RPP-08, Lee-RPP-08, Norman-AP-05}.
  A variety of models have been proposed to describe this phase.
   At present a consensus is lacking and the merits of the different
   models are being vigorously debated.
   Some models propose the partial truncation of the Fermi surface in
   the pseudogap phase is due to the presence of order, which breaks
   translational symmetry. However in the absence of experimental evidence
   for broken translational symmetry \cite{Tomeno-PRB-94}, models without this property
   have gained traction. These in turn can be divided into two classes.
   One emphasizes the reduction of the superconducting (SC) $T_{c}$ by
   strong phase fluctuations due to the reduced superfluid density in UD cuprates.
   This allows the larger SC energy gaps, which in the d-wave form are maximal
   in the antinodal directions on the Fermi surface, to remain finite
   at temperatures $T>T_{c}$ due to preformed Cooper pairs
   but without off-diagonal long order \cite{Norman-PRB-07}.
   Nernst effect and diamagnetism experiments confirm the presence of SC fluctuations
   in an extended temperature range above $T_{c}$ in the UD region of the phase diagram
   but the range ends substantially below the temperature scale of the
   onset of pseudogap behavior \cite{Yayu-PRB-06, Lilu-PRB-10}.
   Alternative models interpret the anomalous properties of the pseudogap phase
   as precursors to the Mott insulator at zero doping \cite{Lee-Wen-RMP}.
Since Mott insulating behavior \textit{per se} is not associated
with translational symmetry breaking, precursor behavior may not
be associated with it either.

Andreev tunneling has been proposed as a distinguished tool to discriminate
  between SC pairing fluctuations and precursor insulating in the pseudogap phase
in an early paper by Deutscher \cite{Deutscher-nature}. (For a review of Andreev reflection of pure d-wave superconductor see Ref\cite{Tanaka} ) It was pointed out that
 the voltage (or energy) scale in Andreev tunneling experiments
  on UD cuprates in the pseudogap phase \cite{Yagil-PhysC-95}
  was substantially below that observed in Giaever or single particle
  tunneling experiments. The two voltage scales however were the same
  in the overdoped (OD) region. In a subsequent review of the Andreev experiments Deutscher
concluded that ``the balance was tilted somewhat against the
preformed pairs scenario'' \cite{Deutscher-RMP}. However, the opposite conclusion,
namely that higher pseudogap energy scale reflects the pairing
strength while a second lower scale the SC condensation energy, was argued for in a later review by Hufner and coworkers which examined many experimental results
using different techniques\cite{Hufner-RPP-08}. To make progress in this debate one needs to move beyond qualitative arguments about many individual experiments and on to more explicit models which can be used to consistently analyze a whole
range of experiments.

In this letter, we perform such solid analysis by using the model proposed by Yang, Rice and Zhang (YRZ), which has successfully been applied to explain many other experiments, to study both Andreev
reflection and Giaever type tunneling for the UD cuprates. Good agreements are achieved, with
both the Andreev reflection where a small SC gap $\Delta_{s}$ is reported, and
Giaever type tunneling experiment where a large pseudogap is
reported. Our theory provides a semi-quantitative description of the
two gap scenario.

\begin{figure}[t]
\centerline
{
\includegraphics[width = 4.0cm, height =10.5cm, angle= 270]
{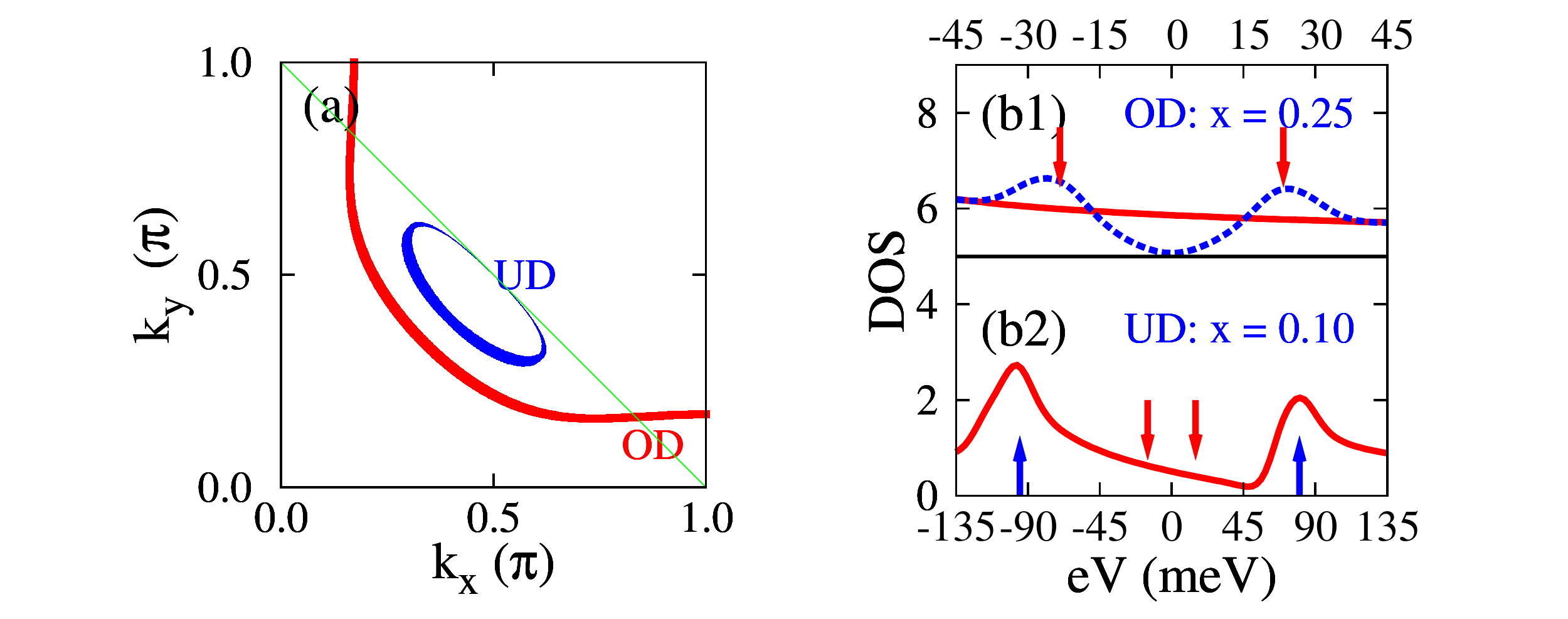}
}
 \caption[]
{ Panel (a): Fermi surface in YRZ model: Blue pocket for
UD ($ x= 0.10$) and red curve for OD ($x=
0.25$). The width on the pocket (blue) represents spectral weight of
quasi-particle. Plane (b): DOS in the model for UD
and for OD, with the gaps denoted by arrows.} \label{fig:YRZ_FS}
\end{figure}

In the YRZ model, a single particle propagator was proposed for
the pseudogap phase\cite{YRZ}. The YRZ propagator was inspired by
an analysis by Konik, Rice and Tsvelik \cite{Tsvelik-ladder} of a
2D array of lightly doped 2-leg Hubbard ladders which gave a set
of hole pockets with an energy gap on fixed lines in
$\boldsymbol{k}$-space connected by elastic Umklapp scattering
processes. The adaptation of this analysis to a 2D Hubbard model
for a square lattice lightly doped away from half-filling was
influenced by the functional renormalization group results at weak
to moderate interaction strength of Honerkamp, Salmhofer and
collaborators \cite{Honerkamp-PRB-01} and by Zhang \textit{et al}
\cite{RMFT} early analysis of Anderson's resonant valence bond
(RVB) proposal \cite{Anderson-RVB, Plain-vanilla}. The strong
coupling t-J model was analyzed using a Gutzwiller renormalized mean field
theory (RMFT). In this phenomenological approach the single
particle gap at the antinodes is controlled by the RVB gap which
truncates the tight-binding Fermi surface into 4 pockets centered on
the nodal directions (see Fig[\ref{fig:YRZ_FS}a]). The d-wave SC energy gap
opens up primarily on these Fermi pockets.

This phenomenological propagator gives a consistent
description~\cite{YRZ} of angle resolved photoemission spectroscopy
(ARPES) experiments which followed the evolution of the Fermi
surface from 4-disconnected Fermi arcs centered on the nodal
directions in UD to the full Fermi surface in OD cuprates \cite{Damascelli-RMP}. The model has  been used to
explain  a range of other spectroscopic measurements
\cite{Yang-EPL-09}, e.g. ARPES results showing increasing
particle-hole asymmetry as one moves away from the nodal
directions \cite{Yang-nature-08}, angle integrated photoemission
electron spectroscopy (AIPES) experiments measuring the doping,
$x$ , dependence of the density of states (DOS)
\cite{AIPES-PRB-09} and scanning tunneling microscopy (STM)
measurements of the coherent Bogoliubov quasiparticle dispersion
at low temperatures \cite{Kohsaka-Nature-08}. The YRZ form was
also used by Carbotte, Nicol and coworkers to successfully
describe the $T$ and $x$ dependences of a wide range of properties
in the pseudogap phase including specific heat
\cite{LeBlanc-PRB-specificheat}, optical
conductivity\cite{Illes-PRB-optical}, London penetration depth
\cite{Carbotte-PRB-Penetration} and symmetry dependent Raman
scattering spectra \cite{LeBlanc-PRB-Raman}. The latter was also
analyzed in a similar way by Valenzuela and Bascones
\cite{Leni-PRL-07}.

In the YRZ model \cite{YRZ}, the incoherent part of the single particle Green's function
contributes a smooth spectral background with a tiny real component at low energies, and the coherent part reads,
\begin{eqnarray}
G^{0}{(\boldsymbol{k}, \omega)} ={g_{t}} / {\big[ \omega - \epsilon_{\boldsymbol{k}} -
\Sigma^{0}{(\boldsymbol{k},\omega)} \big]}
\label{eq:RVB}
\end{eqnarray}
where $g_{t}= 2x / (1+x)$ is a renormalization factor \cite{RMFT,
Plain-vanilla}. We use the result from the RMFT for the RVB state
as the ``bare" dispersion $\epsilon_{\boldsymbol{k}}$
\cite{dispersion}. $\Sigma^0$ is the self energy, which is zero
for the OD cuprates  ($x > x_c = 0.2$). For the UD cuprates, $x <
x_c$,
$
\Sigma^{0}{(\boldsymbol{k}, \omega)} =
{\big[\Delta^{RVB}_{\boldsymbol{k}} \big]^{2}} / {\big( \omega +
\epsilon^{0}_{\boldsymbol{k}} \big)}
$with $\epsilon^{0}_{\boldsymbol{k}} = -2t (\cos{k_{x}} +
\cos{k_{y}} )$ \cite{dispersion}, and
$\Delta^{RVB}_{\boldsymbol{k}}= \Delta_{0} (1-x/x_{c})(\cos{k_{x}}
- \cos{k_{y}})$, with $ \Delta_{0} = 0.3$. All energies are in
unit of $t_{0} = 0.3eV$. Note that $\epsilon^0_{\mathbf{k}} =0$
at the antiferromagnetic reduced Brillouin zone boundary, where
the Umklapp scattering is strongest \cite{Honerkamp-PRB-01}. Eq[\ref{eq:RVB}] predicts four Fermi pockets in the pseudogap phase, consistent
with the recent laser ARPES data \cite{Zhou-Nature-09}.
Fig[\ref{fig:YRZ_FS}a] shows one of these Fermi pockets.

Here we consider a d-wave
superconducting gap function, $\Delta^{sc}_{\boldsymbol{k}}= \Delta^{sc}_{0}
(\cos k_{x} - \cos k_{y})$ for the states around the Fermi surface
within a small energy shell. Several previous work proposed similar SC pairing form of YRZ model \cite{YRZ,
Carbotte-PRB-Penetration,Yang-EPL-09}. We choose $\Delta^{sc}_{0} = 0.08
t_{0}$ for UD ($x = 0.1$), and $\Delta^{sc}_{0} =
0.04t_{0}$ for OD ($x = 0.25$). These gap parameters lead to the SC gap $\Delta_{s}$ (maximum $ \mid \Delta^{sc}_{\boldsymbol{k}} \mid$ on Fermi surface)
comparable to those observed in the Andreev reflection ($\sim$15
and 21 meV, respectively) \cite{Yagil-PhysC-95}, and to those
reported in the recent STM data \cite{Kohsaka-Nature-08}. The
Green's functions for the SC state in the UD and OD cuprates take
the following form in Nambu spinor representation,
\begin{eqnarray}
&&G^{sc}_{UD}{(\boldsymbol{k}, \omega)}= g_t\sum_{i= 1,2}
Z_{\boldsymbol{k},i} \frac{\omega + E_{\boldsymbol{k},i} \tau_{z}
- \Delta^{sc}_{\boldsymbol{k},i} \tau_{x}}
{\omega^{2} - E_{\boldsymbol{k},i}^{2} - (\Delta^{sc}_{\boldsymbol{k},i})^{2}},  \nonumber \\
&&G^{sc}_{OD}{(\boldsymbol{k},\omega)} = g_t \frac{\omega +
\epsilon_{\boldsymbol{k}} \tau_{z} - \Delta^{sc}_{\boldsymbol{k}}
\tau_{x}} {\omega^{2} - \epsilon_{k}^{2} - (\Delta^{sc}_{k})^{2}},
\label{eq:sc}
\end{eqnarray}
where $\tau_{x/y/z}$ are the Pauli matrices, and the label $i=1,\,
2$ denotes the lower and higher energy quasiparticle bands given
by Eq[\ref{eq:RVB}] for the UD  region with spectral weight
$g_tZ_{\boldsymbol{k}, i}$ and dispersion $E_{\boldsymbol{k},i}$.
$G^{sc}_{OD}$ has a BCS form. In the UD region, the quasiparticle
energy of the band $i=2$ is well above the chemical potential, which supports our choice $\Delta^{sc}_{\boldsymbol{k},2} =0$. The typical
profile of DOS of the SC state is shown in Fig[\ref{fig:YRZ_FS}b].
In the UD case, the peaks at $\pm 90$ meV are related to the RVB
gap around the antinodes.

 \begin{figure}[t]
\centerline
{
\includegraphics[width = 7.0cm, height =4.0cm, angle= 0]
{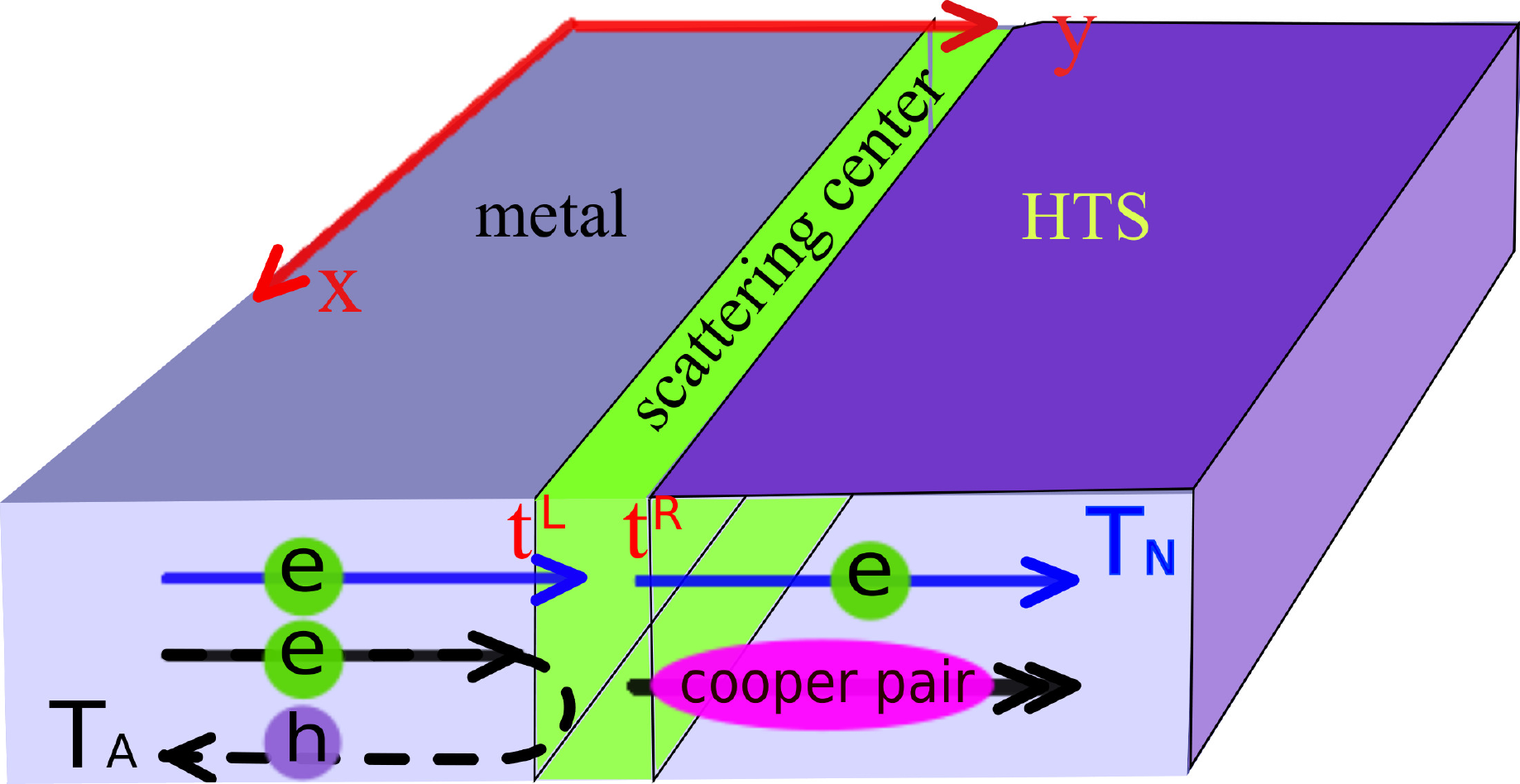}
}
 \caption[]
{Schematic plot for a metal-superconductor junction.  Left: normal
metal; Mid, scattering center, and Right: high Tc Cu-oxide
superconductor. $T_N$ and $T_A$ represent single particle and Andreev
reflection tunnelings, respectively.
 }
\label{fig:NS}
\end{figure}

We use the Keldysh formalism to study Andreev reflection of a NS
junction, consisting of a normal metal lead at the left
and a superconductor lead at the right as shown in
Fig[\ref{fig:NS}]. We approximate the connection area or line
interface as a scattering center. The tunneling of an electron from the
left(L) to the right(R) via the scattering center can be described by the hybridization  Hamiltonian,
$H^{\prime} = \sum_{\boldsymbol{k}, \alpha = L/R}
(t^{\alpha}_{\boldsymbol{k}} c^{\dag}_{\boldsymbol{k}, \alpha} d +
h.c.) $ with $c_{\boldsymbol{k},\alpha}$ and $d$ the electron
annihilation operators at lead $\alpha=L, \, R$ and on scattering
center, respectively. In our following calculations, we assume
$t^{\alpha}_{\boldsymbol{k}}= t^{\alpha}$ for simplicity. In an
Andreev reflection process (for a review see
Ref\cite{Tinkham-PRB-82}), an electron in the normal metal moving
along the $y$-direction to the scattering center may be reflected
back as a hole due to the proximal
superconductivity in the scattering center.
In the tunneling process, $k_x$ is conserved.
 As studied by Meir \textit{et al.} \cite{Meir-PRL-92} for NN junction, and by Sun \textit{et al.} \cite{sun-PRB-99} and Wang \textit{et al.} \cite{Wang-PRL-09} for NS junction,
 the total current can be given by  (setting $\hbar=1$),
\begin{eqnarray}
I= \sum_{s=\pm}es \int d\omega \frac{dk_{x}}{2\pi}
 \big \{ T_{N}^{ss}(k_{x},\omega) [f^{L}(\omega +s eV) - f^{R}(\omega) ] \notag \\
 + T_{A}(k_{x},\omega) [f^{L}(\omega + seV) - f^{L}(\omega - seV) ] \big \},
\end{eqnarray}
where $f^{\alpha}$ is the Fermi function at lead $\alpha$.
$T^{ss}_{N}, \, T_A$ are the diagonal components of the single
particle tunneling matrix $T_{N}$ and the Andreev reflection
coefficient, respectively. $s = + (-)$ corresponds to the electron
(hole) channel. With the frequency-dependence implicitly implied
we have,
\begin{eqnarray}
&&T_{N} (k_{x})  = G^{a}_{c}(k_{x}) \Gamma^{L} (k_{x})G^{r}_{c} (k_{x}) \Gamma^{R} (k_{x}) \\
&&T_{A}(k_{x}) = [G^{a}_{c}(k_{x})]_{12} [ \Gamma^{L} (k_{x})] _{22} [G^{r}_{c} (k_{x}) ]_{21}
[\Gamma^{L} (k_{x})]_{11} \notag
\end{eqnarray}
where $G^{a/r}_c$ refers to the retarded or advanced Green's
function, $G_{c}(k_{x}) =1/ \big[ \omega - \Sigma^{L} (k_{x})
-\Sigma^{R} (k_{x})  \big]$ is the Green's function in the
scattering center \cite{Giaever-scattering}. The self energies $\Sigma^{\alpha} (k_{x}) =
(t^{\alpha})^{2} \tau_{z} G^{\alpha} (k_{x}) \tau_{z}$ and lifetime broadening $\Gamma^{\alpha}(k_{x}) =  -i
\big[\Sigma^{\alpha, a} (k_{x}) - \Sigma^{\alpha,r} (k_{x})\big]$,
with $G^{\alpha}(k_{x})$ the Green's functions on the metal side or on the superconductor side given by
Eq[\ref{eq:RVB},\ref{eq:sc}] at their edges to the scattering
center. Spin rotational invariance leads to $T^{++}_{N}(\omega) =
T^{--}_{N}(-\omega)$ and $T_{A}(\omega) = T_{A}(-\omega)$.

The conductance at zero temperature is given by
\begin{eqnarray}
 \sigma_{s}(eV) = \int_{-\pi}^{\pi} \frac{d k_{x}} {2\pi} e^{2}\sum_{s =
\pm}\big[ T_{N}^{ss}(k_{x}, seV)
 + 2 T_{A}(k_{x}, seV) \big].
 \label{eq:conductance}
\end{eqnarray}
In our calculations, we consider the tunneling along an antinodal
direction of the CuO$_2$ plane, and assume a parabolic dispersion
on the normal metal, $\epsilon^{L}_{\boldsymbol{k}} = {(k^{2}
-k_{F}^{2})}/{2m}$, setting $k_{F} = \pi/2$, $m = \pi/2$ in
unit of $1/t_{0}$.

\begin{figure}[t]
\centerline
{
\includegraphics[width = 4cm, height =9.0cm, angle= 270]
{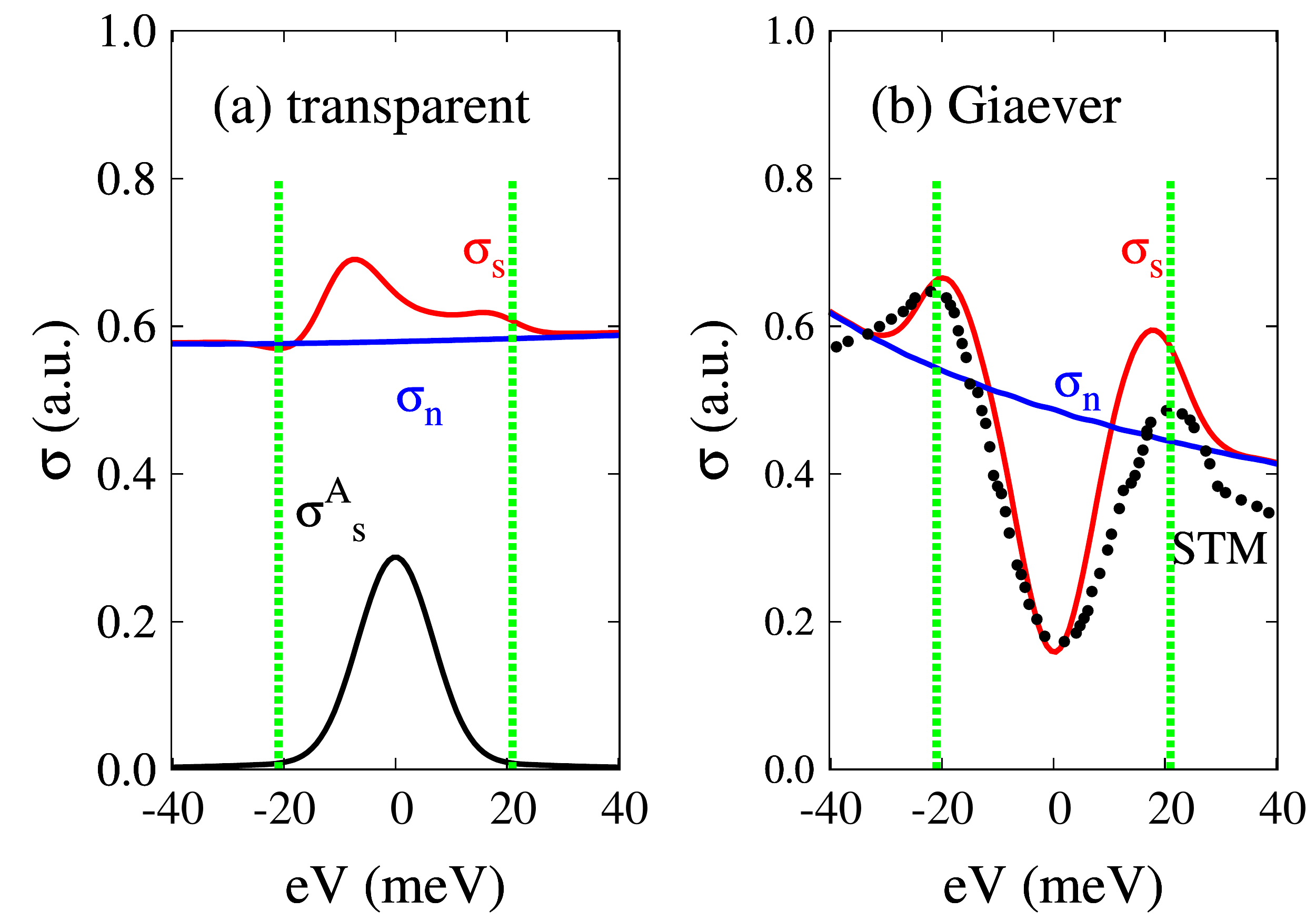}
}
 \caption[]
{Tunneling conductance for OD cuprate ($x= 0.25$). Panel (a) transparent limit. Panel (b) Giaever limit.
$\sigma_{s}$ ($\sigma_{n}$) for NS (NN) junction.
$\sigma_s^{A}$ is the contribution to $\sigma_{s}$ from Andreev reflection. The SC
gap ($\Delta_{s}\sim 21$ meV) is indicated with green lines. Black dots in Panel (b) is STM $dI/dV$ for overdoped Bi2212 (68K) \cite{Yazidani-science-08}.
 (broadening
$\delta = 0.005$. $t^{L} = g_{t}^{1/2}t^{R} = 10 \mbox{, }
0.02$ for the two limits, respectively.) } \label{fig:OD}
\end{figure}

We now discuss the tunneling conductances in our model and compare
them with experiment.  There are two parameter regions in  the
tunneling. At large $t^R/L$, we have transparent tunneling,
Andreev reflection is dominant within the SC gap $\Delta_{s}$. At small $t^R/L$,
the Andreev reflection is strongly suppressed, and single particle
tunneling dominates.

We first
discuss the OD case, which is similar to the conventional BCS
superconductor. In this case, the normal state is a metal
with a full Fermi surface shown in Fig[\ref{fig:YRZ_FS}a]. The RVB
gap vanishes, leaving only a SC gap $\Delta_{s}$.
 The calculated conductance $\sigma_s$
in a typical transparent region is shown in Fig[\ref{fig:OD}a].
The conductance in the single particle tunneling region is shown
in Fig[\ref{fig:OD}b] in a good agreement with STM data \cite{Yazidani-science-08}, where the contribution from the Andreev
reflection is essentially vanishing. 

\begin{figure}[t]
\centerline
{
\includegraphics[width = 4cm, height =9.0cm, angle= 270]
{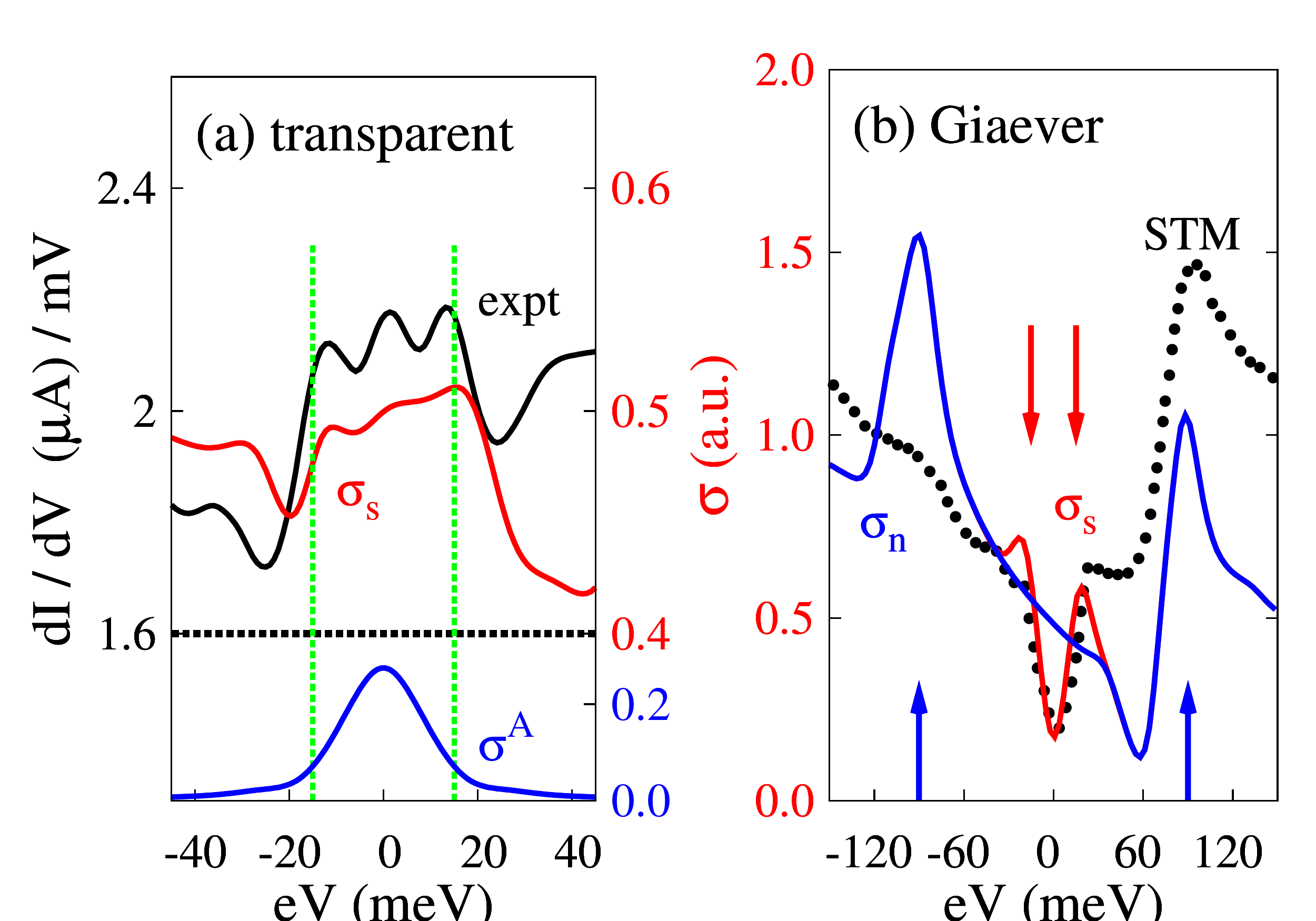}
}
 \caption[]
{Conductance for UD cuprate ($x = 0.1$). Panel (a)  transparent
limit. Red curve is the NS conductance $\sigma_{s}$,
$\sigma^{A}_{s}$ in blue denotes the Andreev reflection
contribution, black curve is experimental Andreev reflection data
on UD YBCO samples \cite{Yagil-PhysC-95}. The SC gap ($\Delta_{s}\sim$ 15 meV) is indicated
with green lines. Panel (b) Giaever tunneling limit. Blue
curve is $\sigma_{n}$ for NN junction. Black curve is $dI/dV$
curve for UD Bi-2212 (T$_{c} = 45$ K) observed in STM
\cite{Kohsaka-science-07}. The two energy gaps are marked by
arrows. The same parameters as OD shown in Fig[\ref{fig:OD}] are used. }  \label{fig:UD}
\end{figure}

For the UD cuprates, there are two energy scales, a larger RVB gap
or pseudogap, and a smaller SC gap $\Delta_{s}$.  We expect two distinct energy
scales in the tunneling conductance. $\sigma_s$ in the transparent
region is plotted in Fig[\ref{fig:UD}a]. The contribution from the
Andreev reflection is only substantial within the SC gap on the
remnant region of the Fermi surface. $\sigma_s$ shows a clear
peak-edge feature at the SC gap energy $\Delta_{s}$. Our result is in good
agreement with the reported Andreev tunneling experiment
\cite{Yagil-PhysC-95}, which is reproduced here for comparison. In
a single gap scenario, it is not clear that the relative
contributions of Andreev and single particle processes to
tunneling, into the ordered SC state should change between OD and
UD.

The voltage-dependent conductance in the single particle tunneling
region is plotted in Fig[\ref{fig:UD}b] for UD samples. In this case, the Andreev
reflection is substantially suppressed, so that
voltage-dependence of $\sigma_s$ is similar to that of DOS
(Fig[\ref{fig:YRZ_FS}b]). In both $\sigma_s$ and DOS in the SC
phase, there are two energy scales. The lower energy peak is
associated with the SC gap $\Delta_{s}$, and the higher energy peak with the
RVB gap.  The overall profile we obtained is very close to the recent
STM experimental data on Bi-2212 \cite{Kohsaka-science-07}. Note
that the tunneling conductance we study here does not include a
strong electron-hole asymmetry in the spectral weight associated
with the asymmetry of injecting an electron or a hole in the
strongly correlated systems, as examined by Anderson and Ong
~\cite{Anderson-Ong} and by Randeria \textit{et al.}
\cite{RMFT-asymmetry}. We argue that the relatively low energy
tunneling spectral is weakly affected by this strong asymmetry
\cite{RMFT-asymmetry} so that our model calculations of the
tunneling conductance remain valid in the relevant energy region.

  One point worth noting concerns the asymmetry in the single particle
  tunneling spectra between positive and negative voltages.
  This is much more pronounced in the DOS calculations based on
  the YRZ propagator than in the tunneling experiments.
  The asymmetry in the YRZ DOS comes from the Dirac point
  at positive energies in the lower quasiparticle band.
  Interestingly a linear dependence of the DOS at the chemical
potential on the hole density was reported in AIPES \cite{AIPES-PRB-09},
which is consistent with the approaching to a Dirac point as the hole density is reduced.
The quasiparticle DOS calculations do not include the effects of lifetime broadening
of the coherent quasiparticle spectra.
This could be important in view of the evidence
for strong inelastic scattering processes.
These could act to smear out signs of a Dirac point
at finite energies above the chemical potential and
so reduce the asymmetry in the DOS. This discrepancy requires further study.

In summary, we have theoretically studied the tunneling
spectroscopy of the underdoped cuprates.  The single particle
pseudogap in our theory appears as a precursor to the Mott
insulator at zero doping. Our theory semi-quantitatively explains
the two gap scenario at underdoping revealed in tunneling
experiments: a small energy gap associated with Andreev reflection
in the transparent limit and a large gap associated with single
particle tunneling, which are in contrast with the same scales at
overdoping \cite{Deutscher-nature, Deutscher-RMP}. Although the
onset of long range superconducting order on the remaining Fermi
pockets (or arcs) induces a reduced pairing amplitude also in
these antinodal regions through Cooper channel scattering
processes, the corresponding Andreev signal is found strongly
suppressed. Since the tunneling experiments, \textit{per se}, do
not address the origin of this two gap behavior, from broader
viewpoint, our calculations support all models in which the pseudogap
at underdoping is due to the partial truncation of the Fermi
surface through an insulating gap in the antinodal regions, but
not due to preformed Cooper pairs.

We thank Manfred Sigrist and Qiang-Hua Wang for discussions. Supports from Swiss National funds and the NCCR MaNEP (K.Y. Y. and T. M. R.), and RGC grant of HKSAR (K. H., W. Q. C. and F. C. Z.) are also gratefully acknowledged. T. M. R. was supported in part by the Center for Emergent Superconductivity, an Energy Frontier Research Center supported by the US DOE, Office of Basic Energy Sciences, and by a Visiting Professorship at the University of Hong Kong.


\begin{thebibliography} {}
\bibitem{dispersion}
We use a "bare" dispersion from the RMFT for the t-J model,
 $\epsilon_{\boldsymbol{k}} = -2t (\cos{k_{x}} + \cos{k_{y}} ) -
4g_{t}t^{\prime} \cos{k_{x}} \cos{k_{y}} - 2g_{t}t^{\prime\prime}
(\cos{2k_{x}} + \cos{2k_{y}}) - \mu$ with parameters $t = (g_{t}
t_{0} + \frac{3}{8} g_{s}J \chi)$, $t^{\prime} / t_{0} = -0.4$,
$t^{\prime \prime} /t_{0}= 0.2$, $J/t_{0} = 0.3 $, $\chi_{0}
=0.338$, $g_{s} = 4/(1+x)^{2}$. $t_0$ is about $0.3$ eV.
\bibitem{Giaever-scattering}
 {In Giaever limit, the scattering center has a uniform tunneling channel at all energies in our calculation.}
\end{thebibliography}
\end{document}